\documentclass[review]{elsarticle}

\usepackage{lineno,hyperref}
\modulolinenumbers[5]

\usepackage{graphicx}
\usepackage{dcolumn}
\usepackage{bm}
\usepackage{amsmath}

\journal{Journal of \LaTeX\ Templates}

\biboptions{authoryear}

\begin{document}

\begin{frontmatter}

\title{Mass-imbalanced atoms in a hard-wall trap: an exactly solvable model associated with $D_{6}$ symmetry}

\author{Yanxia Liu\fnref{bei,shan}}

\author{Fan Qi\fnref{shan}}

\author{Yunbo Zhang\fnref{shan,lead}\corref{corres}}
\ead{ybzhang@sxu.edu.cn}

\author{Shu Chen\fnref{bei,bei1,li}\corref{corres}}
\ead{schen@iphy.ac.cn}

\fntext[bei]{Beijing National Laboratory for Condensed Matter Physics, Institute of
Physics, Chinese Academy of Sciences, Beijing 100190, China}
\fntext[shan]{Institute of Theoretical Physics, Shanxi University, Taiyuan 030006, P. R.
China}
\fntext[bei1]{School of Physical Sciences, University of Chinese Academy of Sciences,
Beijing 100049, China}
\fntext[li]{The Yangtze River Delta Physics Research Center, Liyang, Jiangsu 213300,
China}
\fntext[lead]{Lead Contact}
\cortext[corres]{Correspondence:}

\end{frontmatter}

\section*{SUMMARY}
We show that a system consisting of two interacting particles with mass ratio $3$ or $1/3$ 
in a hard-wall box can be exactly solved by using Bethe-type ansatz. The ansatz is based on a finite
superposition of plane waves associated with a dihedral group $D_{6}$,
which enforces the momentums after a series of scattering and reflection
processes to fulfill the $D_{6}$ symmetry. Starting from a two-body elastic
collision model in a hard-wall box, we demonstrate how a finite momentum
distribution is related to the $D_{2n}$ symmetry for permitted mass ratios. For
a quantum system with mass ratio $3$,
we obtain exact eigenenergies and eigenstates by solving Bethe-type-ansatz
equations for arbitrary interaction strength. A many-body excited state of
the system is found to be independent of the interaction strength, i.e. the
wave function looks exactly the same for non-interacting two particles or in
the hard-core limit.

\section*{INTRODUCTION}

%Symmetry has always been one of source of obsession in physics, from the Lorentz invariance of electrodynamics,  which Albert Einstein made into the cornerstrone and unifying principle of an entirely new understanding of space and time, to the gauge symmetries underlying the standard model of strong and electroweak interactions \citep{Mottola,Gross}. In classical and quantum mechanics, symmetry was used as a way to derive new solution and reveal the physical nature. The most common are the translation, discrete translation reflection and scaling symmetry. This article will show that for mass-imbalanced system, the integrable point is associated to the finite group. The symmetry play important roles to describe the solvable quantum particle systems.

Exactly solvable models have played an important role in the understanding
of the complexity of interacting quantum systems, especially in one
dimension \citep{Albeverio,Sutherland,Gaudin2014,Takahashi,Gutkin}. Prominent
examples include the Lieb-Liniger model for interacting bosons \citep{Lieb},
the Gaudin-Yang model for two-component fermions \citep{Yang1967}, and the
extended family of multi-component Calogero-Sutherland-Moser (CSM) models
\citep{Sutherland1968}. These models provide ways to exploring and
understanding the physics of quantum few-body and many-body systems. An
elegant example of solvable few-body models is the system of two interacting
atoms in a harmonic trap \citep{Busch1998}, which has become a benchmark in
the exploration of interacting few-body system, even in the accuracy
estimate of numerical procedure for interacting few particles.

Experiments with few cold atoms provide unprecedented control on both the
atom number $N$ with unit precision and the interatomic interaction strength
by combination of sweeping a magnetic offset field and the confinement
induced resonance \citep{Chin}. The experiments have so far realized the
deterministic loading of certain number of atoms in the ground state of a
potential well \citep{Serwane}, the controlled single atom and atom pair
tunneling out of the metastable trap \citep{Zurn2012,Zurn2013}, the
preparation of quantum state for two fermionic atoms in an isolated
double-well \citep{Murmann}, etc. The crossover from few- to many-body
physics has been shown by observing the formation of a Fermi sea one atom at
a time \citep{Wenz}. In the strongly interacting limit an effective
Heisenberg spin chain consisting of up to four atoms can be
deterministically prepared in a one-dimensional trap \citep{Murmann2015}.

While most of the exactly solvable interacting models are limited to the
equal-mass case, recently much attention has been drawn on one-dimensional
(1D) mass-imbalance systems composed of hard-core particles
\citep{Olshanii2015,Harshman,Scoquart,Olshanii2018,Dehkharghani,Volosniev}. It is
found that some few-body systems are solvable if the hard-core particles
with certain masses are arranged in a certain order. A quantum four-body
problem associated with the symmetries of an octacube is exactly solved for
hard-core particles with specific mass ratio and its exact spectrum stands
in good agreement with the approximate Weyl's law prediction \citep{Olshanii2015}.
In a Bose-Fermi superfluid mixture, especially of two
mass-imbalance species, macroscopic quantum phenomena are particularly rich
due to the interplay between the Bose and Fermi superfluidity \citep
{Salomon,Yao}. Different from the integrable systems with their
integrability guaranteed by the existence of Yang-Baxter equation and a
series of conserved quantities \citep{Gaudin1971,McGuire,GuanXW,HaoYJ},
reliable criteria for the solvability of mass-imbalanced systems are still
lack.

For an interacting system with a finite interaction strength, the
mass-imbalance system is generally not exactly solvable \citep%
{Deuretzbacher,Pecak,Pecak2}. Particularly, when the system is in an
external trap, the interacting problem with different masses becomes
complicated and it is hard to get an analytical solution even for a
two-particle system since the external potential brings about the coupling
of center-of-mass and relative coordinates \citep{Deuretzbacher,ChenXing} and
generally one can not completely separate the relative motion of particles
from the others. In this work we study the mass-imbalanced two-particle
system with finite interaction strength in a hard-wall trap and give the
Bethe-type-ansatz solution of the system with mass ratios $3$ or $1/3$. The Bethe-type ansatz is based on finite
superpositions of plane waves, which is generally not fulfilled for the
mass-imbalance system as each collision process generates a new set of
momentums. When the
mass ratio takes some special values, we find that the motion of classical particles after multiple collisions can be
characterized by finite sets of momentums, which is associated with  the nonergodicity condition \citep%
{Richens,Evans,nonergodicity1,nonergodicity2} of the classical elastic
collisions of particles with different masses in the hard-wall box. When the mass ratios are at these nonergodicity points, it is interesting to find that
the permitted momentums of particles fulfill the symmetry described by the dihedral group $D_{2n}$.The existence of finite momentums
enables us to take the wavefunction of
the two-body quantum system as Bethe-type ansatz, i.e., as the superposition of all plane waves with permitted momentums. While the equal-mass case corresponds to the solvable Lieb-Liniger model under the open boundary condition,  we find that
only the mass-imbalance case with mass ratios $3$ or $1/3$ is exactly solvable, i.e.,
only the case with quasimomentums fulfilling the $D_{6}$ symmetry is exactly solvable.

The paper is organized as follows. In section II, we first discuss the
nonergodicity condition for the classical collision problem in a hard-wall
box and show how the momentums with specific mass ratios are related to
the $D_{2n}$ symmetry. In section III, we study the quantum system with mass
ratio $\eta =3$ by using the Bethe-type-ansatz wavefunction, which permits us to
get the Bethe-type-ansatz equations for all interaction strengths. Solving the
Bethe-type-ansatz equations, we can get the quasimomemtum distribution of the
system and thus the exact eigenstates and eigenvalues. A summary is given in
the last section.

\section*{NONERGODICITY CONDITION FOR COLLISION IN A HARD-WALL TRAP}

First we consider a classical collision problem of two particles with
unequal masses $m_{1}$ and $m_{2}$ in a one-dimensional hard-wall trap.
There are two types of collision processes, namely, the scattering between
particles and the reflection process when the particle hits the wall. Write
the momentums of two particles before and after the collision as vectors as $%
\mathbf{k}=(k_{1}, k_{2})^T$ and $\mathbf{k^{\prime }}=(k_{1}^{\prime },
k_{2}^{\prime })^T$, respectively. For the elastic scattering in which both
total momentum and total energy of the particles are conserved, we have
\begin{eqnarray}
k_{1}+k_{2} &=&k_{1}^{\prime }+k_{2}^{\prime },  \label{m1} \\
\frac{k_{1}^{2}}{2m_{1}}+\frac{k_{2}^{2}}{2m_{2}} &=&\frac{{k_{1}^{\prime }}%
^{2}}{2m_{1}}+\frac{{k_{2}^{\prime }}^{2}}{2m_{2}}.  \label{energy}
\end{eqnarray}%
From Eq.(\ref{energy}), it is easy to get
\begin{equation}
\frac{k_{1}^{2}-{k_{1}^{\prime }}^{2}}{m_{1}}=\frac{{k_{2}^{\prime }}%
^{2}-k_{2}^{2}}{m_{2}}.
\end{equation}%
Taking advantage of Eq.(\ref{m1}), we see
\begin{equation}
\frac{k_{1}+{k_{1}^{\prime }}}{k_{2}+{k_{2}^{\prime }}}=\eta ,  \label{m2}
\end{equation}%
where the mass ratio $\eta =m_{1}/m_{2}$. By using Eq.(\ref{m1}) and (\ref%
{m2}), it is straightforward to obtain the momentum relation for particle
scattering
\begin{equation}
\mathbf{k}^{\prime }=s\mathbf{k},~~~~\text{with}~~~~s\left( \eta \right)
=\left(
\begin{array}{cc}
\frac{\eta -1}{\eta +1} & \frac{2\eta }{\eta +1} \\
\frac{2}{\eta +1} & \frac{1-\eta }{\eta +1}%
\end{array}%
\right).
\end{equation}%
Here $s$ is an involutory matrix, which satisfies
\begin{eqnarray}
s\left( \eta \right) ^{2}&=&1, \\
s\left( 1/\eta \right) &=&\sigma _{x}s\left( \eta \right) \sigma _{x},
\end{eqnarray}
where $\sigma _{x,y,z}$ are the Pauli matrices. In the case of reflection,
one of the particles changes its sign of momentum. The momentum relation for
reflection is
\begin{equation}
\mathbf{k}^{\prime }=\pm \sigma _{z}\mathbf{k},~~~\text{with}~~~\sigma
_{z}=\left(
\begin{array}{cc}
1 & 0 \\
0 & -1%
\end{array}%
\right) ,
\end{equation}%
where the reflection matrix $\sigma_{z} $ reflects $k_2$ and $-\sigma _{z}$
reflects $k_1$. Notice that the scattering and reflection always occur
alternately and the momentum vector after multiple collisions is
straightforwardly given by successive application of the scattering matrix $%
s $ and reflection matrices $\pm \sigma _{z}$ onto the initial vector, e.g.,
\begin{equation}
\mathbf{k}^{\prime } = s\left( \eta \right) (-\sigma _{z}) s\left( \eta
\right) \sigma _{z} s\left( \eta \right) \sigma _{z}\mathbf{k}
\end{equation}
represents the final momentum vector after three pairs of
scattering-reflection processes, one after another.

\begin{figure}[tbp]
\includegraphics[width=0.75\textwidth]{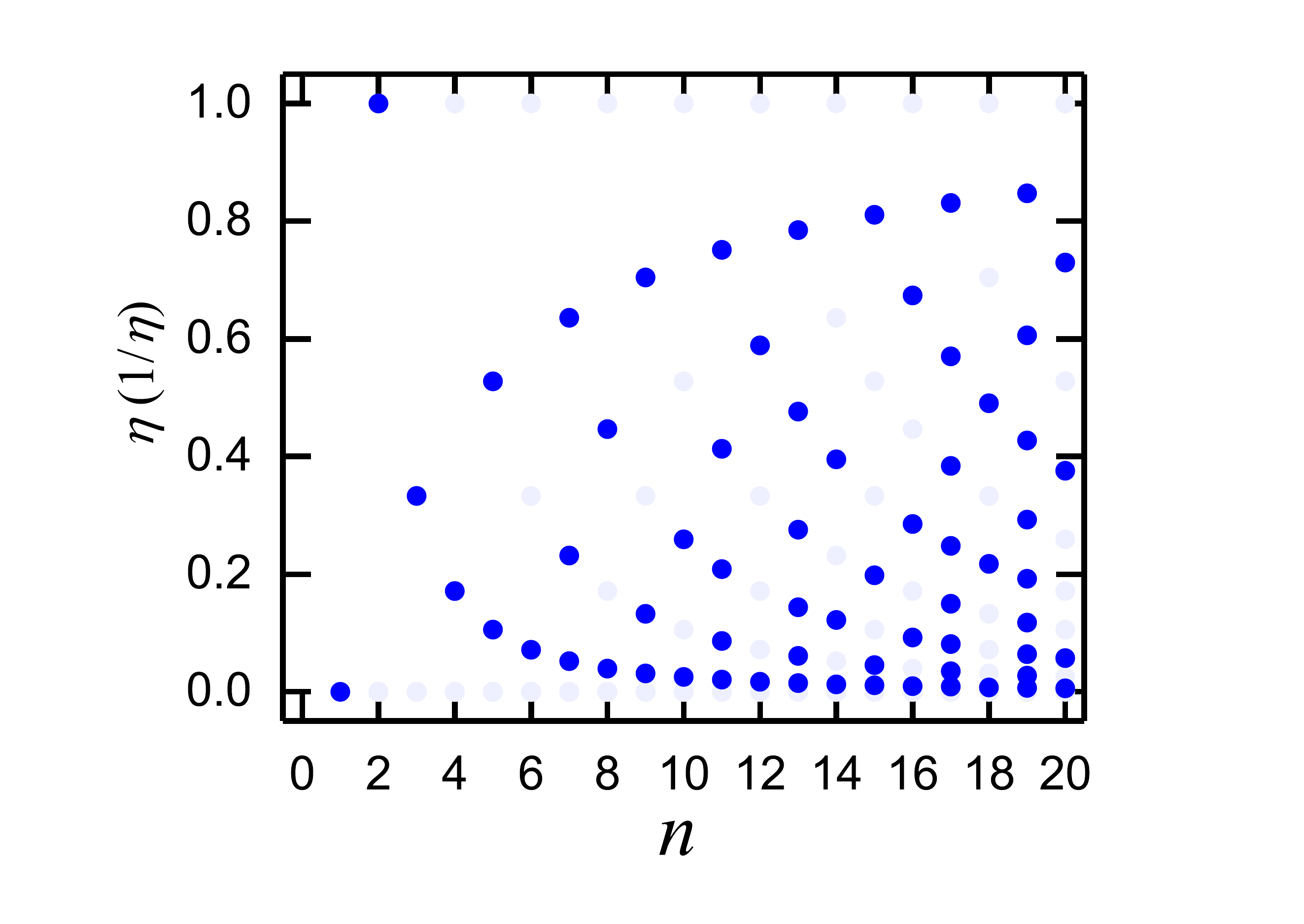}
\caption{The Relation between the Mass Ratio $\protect\eta $ (or $1/\protect%
\eta $) and the Collision Times $n$: $\protect\eta =\tan ^{2}l\protect\pi /2n
$ for Finite Distribution of Momentums} 
{The blue dots represent the solution
correspond to the minimum collision times $n$ and pale blue dots represent
the repeated solutions. There exists a duality for mass ratio $\protect\eta$
and $1/ \protect\eta$. }
\label{fig1}
\end{figure}

Given the initial momentum vector, now we explore how many new vectors may
come into being after multiple collisions. The reflection matrix allows us
to consider only the positive values of the momentum since in the last step
one can always invert the sign by applying either $\sigma_z$ or $-\sigma_z$.
%Although the reflection cannot produce new set of momentum solely, in
%combination with the scattering, the multiple collision processes can lead to various new
%set of momentum. Thus the new set of momentum means that the absolute values of final
%momentum are different from the initial ones.
%Different mass ratio $\eta $ generates very different
%distribution of the momentum after multiple collisions.
In order to effectively study the motion characteristics we may
intentionally structure the collision trajectory such that new momentum
vector would appear in every pair of scattering-reflection processes. There
exist basically two types of trajectories with final momentum vectors
expressed as $\left( -1\right) ^{m}\left(s\sigma _{z}\right) ^{n}\mathbf{k}$
or $\left( -1\right) ^{m}\left( \sigma _{z}s\right) ^{n}\mathbf{k}$. In the $%
n$ pairs of scattering-reflection processes, there are $m$ times reflection $%
-\sigma_z$ and $n-m$ times reflection $\sigma_z$. Usually, the momentum
distribution after multiple collisions becomes rather unpredictable for an
arbitrary mass ratio. However, for some special mass ratios, it is possible
that after multiple collisions the momentum vector will go back to the
initial value. Thus a finite number of momentum vectors form a closed set
with the corresponding collision trajectory being a closed loop, which is
similar to the fixed point in the regular and chaotic motion of particles
bouncing inside a curve \citep{Berry,Sinai}. This means that
\begin{equation}
\left( s\left( \eta \right) \sigma _{z}\right) ^{n}\left(
\begin{array}{c}
k_{1} \\
k_{2}%
\end{array}%
\right) =\pm \left(
\begin{array}{c}
k_{1} \\
k_{2}%
\end{array}%
\right)  \label{C1}
\end{equation}%
or
\begin{equation}
\left( \sigma _{z} s\left( \eta \right)\right) ^{n}\left(
\begin{array}{c}
k_{1} \\
k_{2}%
\end{array}%
\right) =\pm \left(
\begin{array}{c}
k_{1} \\
k_{2}%
\end{array}%
\right),  \label{C2}
\end{equation}%
where $\pm$ corresponds even(odd) $m$ respectively.

After some algebras (see transparent method A for details), we find that the above
equations (\ref{C1}) and (\ref{C2}) are satisfied if the mass ratio $\eta $
and the number of scattering-reflection pairs $n$, hereafter referred to as
collision times, fulfills the following condition
\begin{equation}
\eta =\tan ^{2}l\pi /2n,  \label{nonergodicity}
\end{equation}%
where $l$ and $n$ are positive integers. For given $\eta$, we aim to find
the minimum collision times $n$, after which the momentum sets would be
closed. It suffices that let $l$ be any coprime integer to $n$ and $1\leq
l\leq n$. There exists a duality for mass ratio $\eta$ and $1/\eta$ and in
Fig. 1, we show all qualified mass ratios after $n$-multiple
collisions by blue dots, which are two-fold degenerate for $\eta $ and $%
1/\eta$ except the case of equal mass. A trivial case is that for $\eta=0$
or $1/\eta=0$ which means that there is only one particle left in the
hard-wall trap. The closed set contains but one momentum vector as the only
collision process is the reflection on the left or right wall, which serves
to change it's sign.
%It is not hard to find that the mass ratio which guarantees the nonergodicity
%condition with certain value $n$ automatically do so for collision times
%$n^{\prime }=nl^{\ast }$ where $l^{\ast} $ is the greatest common divisor $l$ and $n$.
%From Fig. 1, we can find that with the increase of the collision times $n$, the
%mass ratio $\eta $ appears repeatedly, e.g. the mass ratio $\eta=3$ is qualified for $n=3,6,9, \cdots$.
%{\bf This is so obvious since first time you pick up $\eta=3$ always means that you will encounter it
%later after another cycle is closed. Is it worth to explain in detail? Or can we just delete the repeated
%blue points in figure 1?}

The equation (\ref{nonergodicity}) assures that for an given initial
momentum vector $\mathbf{k}$ a closed set of finite numbers of the momentum
can be obtained by repeatedly applying the scattering and reflection
operations on it. In the case of equal mass when $\eta =\tan ^{2}\pi /4=1$,
the full momentum set is
\begin{equation*}
\mathbf{k},r\mathbf{k},r^{2}\mathbf{k},r^{3}\mathbf{k},\sigma _{z}\mathbf{k}%
,r\sigma _{z}\mathbf{k},r^{2}\sigma _{z}\mathbf{k},r^{3}\sigma _{z}\mathbf{k}%
,
\end{equation*}%
where $r=s\left( 1\right) \sigma _{z}$ and $r^{2}=-I$ with $I$ the $2\times
2 $ identity matrix. It is easy to find the collision operators $\left\{
I,r,r^{2},r^{3}\right\} $ form a cyclic group $C_{4}$ and $%
\{I,r,r^{2},r^{3},\sigma _{z},r\sigma _{z},r^{2}\sigma _{z},r^{3}\sigma
_{z}\}$ form a dihedral group $D_{4}$. The first nontrivial case arises when
$\eta =\tan ^{2}\pi /3=3$, and the full momentum set consists of
\begin{equation*}
\mathbf{k},r\mathbf{k},r^{2}\mathbf{k},\cdots ,r^{5}\mathbf{k},\sigma _{z}%
\mathbf{k},r\sigma _{z}\mathbf{k},r^{2}\sigma _{z}\mathbf{k},\cdots
,r^{5}\sigma _{z}\mathbf{k},
\end{equation*}%
where $r=s\left( 3\right) \sigma _{z}$ and $r^{3}=-I$. The collision
operators $\left\{ I,r,r^{2},\cdots ,r^{5},\sigma _{z},r\sigma
_{z},r^{2}\sigma _{z},\cdots ,r^{5}\sigma _{z}\right\} $ form a dihedral
group $D_{6}$. Note that when $\eta =\tan ^{2}\pi /6=1/3$, the operators $%
\left\{ I,r,r^{2},\cdots ,r^{5},\sigma _{z},\right.$ %
$\left. r\sigma _{z},r^{2}\sigma_{z},\cdots ,r^{5}\sigma _{z}\right\} $ with $r=-s\left( 1/3\right) \sigma
_{z}$ also form a $D_{6}$ group. This again show the duality for mass ratio $%
\eta $ and $1/\eta $. So we find that Eq.(\ref{nonergodicity}) gives a
series of classical nonergodicity points and the full momentum set can be
written as $\left\{ d_{j}\mathbf{k}|d_{j}\in D_{2n}\right\} $. Here the
dihedral group $D_{2n}$ with $n=2,3,\cdots $ has $4n$ elements, i.e. $%
D_{2n}=\left\{ I,r,r^{2},\cdots ,r^{2n-1},\sigma _{z},r\sigma _{z},\cdots
,r^{2n-1}\sigma _{z}\right\} $, where $r=\pm s\left( \eta \right) \sigma
_{z} $ and $r^{2n}=I$. Here the sign $+$ and $-$ are for $\eta \geq 1$ and $%
\eta<1$, respectively.
\begin{table}[tbp]
\caption{The relationship between mass ratio $\protect\eta $ and the
dihedral group.}
\begin{center}
\begin{tabular*}{0.85\textwidth}{p{0.17\textwidth}<{\centering}p{0.14\textwidth}<{\centering}p{0.14\textwidth}<{\centering}p{0.27\textwidth}<{\centering}cccc}
\hline
$\eta $ & $n$ & $l$ & the dihedral group &  &  &  &  \\ \hline\hline
$0,+\infty$ & $1$ & $1$ & $D_{2}$ &  &  &  &  \\ \hline
$1$ & $2$ & $1$ & $D_{4}$ &  &  &  &  \\ \hline
$1/3$ & $3$ & $1$ & $D_{6}$ &  &  &  &  \\
$3$ & $3$ & $2$ & $D_{6}$ &  &  &  &  \\ \hline
$3-2\sqrt{2}$ & $4$ & $1$ & $D_{8}$ &  &  &  &  \\
$3+2\sqrt{2}$ & $4$ & $3$ & $D_{8}$ &  &  &  &  \\ \hline
$1-2/\sqrt{5}$ & $5$ & $1$ & $D_{10}$ &  &  &  &  \\
$5-2\sqrt{5}$ & $5$ & $2$ & $D_{10}$ &  &  &  &  \\
$1+2/\sqrt{5}$ & $5$ & $3$ & $D_{10}$ &  &  &  &  \\
$5+2\sqrt{5}$ & $5$ & $4$ & $D_{10}$ &  &  &  &  \\ \hline
$7-4\sqrt{3}$ & $6$ & $1$ & $D_{12}$ &  &  &  &  \\
$7+4\sqrt{3}$ & $6$ & $5$ & $D_{12}$ &  &  &  &  \\ \hline
$\vdots$ & $\vdots$ & $\vdots$ & $\vdots$ &  &  &  &  \\ \hline
\end{tabular*}%
\end{center}
\end{table}

In table 1, we list all candidates for the mass ratio $\eta $ which fulfills
the closeness of scattered momentum vector and the corresponding dihedral
group of the collision operators. As $l$ and $n$ are coprime, the value $n$
solely decides the dihedral group $D_{2n}$. For different mass ratio, the
number of momentum vector in the closed set determines the order of the
dihedral group. In Fig. 2, we show the distribution of momentum
in the closed set for $\eta =1$ and $3$, with the emerging $D_{4}$ and $D_{6}
$ symmetry, respectively. Each momentum vector is represented by a point in
the phase space $\left( k_{1},\sqrt{\eta }k_{2}\right) $ where we rescale $%
k_{2}$ by a factor $\sqrt{\eta }$ such that all points are distributed on a
circle due to the energy conservation. It is straightforward to see that the
momentums are distributed on vertices of two 2n-sided polygons, which
fulfill the $D_{2n}$ symmetry. To see it more clearly, we represent the $r$%
-matrix as
\begin{equation}
r=\left(
\begin{array}{cc}
-\cos \frac{l\pi }{n} & \cos \frac{l\pi }{n}-1 \\
\cos \frac{l\pi }{n}+1 \newline
& - \cos \frac{l\pi }{n}%
\end{array}%
\right)
\end{equation}%
which is obtained by inserting the nonergodicity mass ratio\ $\left( \ref%
{nonergodicity}\right) $ into $r= s\left( \eta \right) \sigma _{z}$. Then
performing a similar transformation on $r$, we get%
\begin{equation}
R=UrU^{-1}=\left(
\begin{array}{cc}
\cos \frac{m \pi }{n} & -\sin \frac{m \pi }{n} \\
\sin \frac{m \pi }{n}\newline
& \cos \frac{m \pi }{n}%
\end{array}%
\right) ,
\end{equation}%
where $m=n-l$ and
\begin{equation*}
U=\left(
\begin{array}{cc}
1 & 0 \\
0\newline
& \sqrt{\eta }%
\end{array}%
\right) .
\end{equation*}%
Clearly, $R$ is a two-dimensional rotation matrix, and the standard
presentation of the dihedral group $D_{2n}$ is given by%
\begin{equation}
D_{2n}=\left\{ R,\sigma _{z}|R^{2n}=\sigma _{z}^{2}=1,\sigma _{z}R\sigma
_{z}^{-1}=R^{-1}\right\} .
\end{equation}%
Given an initial set of momentums, the other vertices of polygons are
decided by applying the symmetry operations of the $D_{2n}$ group.

%For $%\eta =1$, as shown in Fig. 2 (a) the permitted momentums are located on the ends of two squares along points $\mathbf{k}\rightarrow r\mathbf{k}\rightarrow r^{2}\mathbf{k}\rightarrow r^{3}\mathbf{k}\rightarrow\mathbf{k}$ and $\sigma _{z}\mathbf{k}\rightarrow r\sigma _{z}\mathbf{k}$ $\rightarrow r^{2}\sigma _{z}\mathbf{k}\rightarrow r^{3}\sigma _{z}\mathbf{k}\rightarrow \sigma _{z}\mathbf{k}$, either of which clearly fulfills the $D_{4}$ symmetry. Similarly, for $\eta =3$, the momentums are located on the vertices of two hexagons as shown in Fig. 2 (b). Generally speaking, for $\eta =\tan ^{2}l\pi /2n$, where $l$ and $n$ are coprime, the momentums are distributed on two regular polygons with $2n$ edges along point $\mathbf{k}\rightarrow r\mathbf{k}\rightarrow r^{2}\mathbf{k}$ $\cdots \rightarrow r^{2n-1}\mathbf{k}$ and $\sigma _{z}\mathbf{k}\rightarrow r\sigma _{z}\mathbf{k}\rightarrow r^{2}\sigma _{z}\mathbf{k}$ $\cdots \rightarrow r^{2n-1}\sigma _{z}\mathbf{k} $, respectively.

%The position of the regular polygons are decided by the initial momentums.

Although the general mass-imbalance collision problem in the hard-wall trap
does not possess discrete symmetries, the momentum distributions in the
phase space exhibit the emergent $D_{2n}$ symmetries in the nonergodicity
points, which includes $2n$ rotational symmetries and $2n $ reflection
symmetries.
%From Fig. 2, we can find that each axis of symmetry joins opposite
%intersection points of the two regular polygons and there are $2n$ symmetry lines.
If the nonergodicity condition (\ref{nonergodicity}) is not fulfilled,
Eqs. (\ref{C1}) and (\ref{C2}) no longer hold true, and the momentum
distribution does not exhibit discrete symmetry. Instead, the momentums
shall distribute on the entire circle with the increase of collision times.

%\emph{For the nonergodicity pionts, when the systems with $\eta =1$, $3$ and $1/3$ are integrable and other cases are pseudointegrable. The paper \citep{Richens} details the reasons why integrable happens in these points.}

\begin{figure}[tbp]
\includegraphics[width=0.75\textwidth]{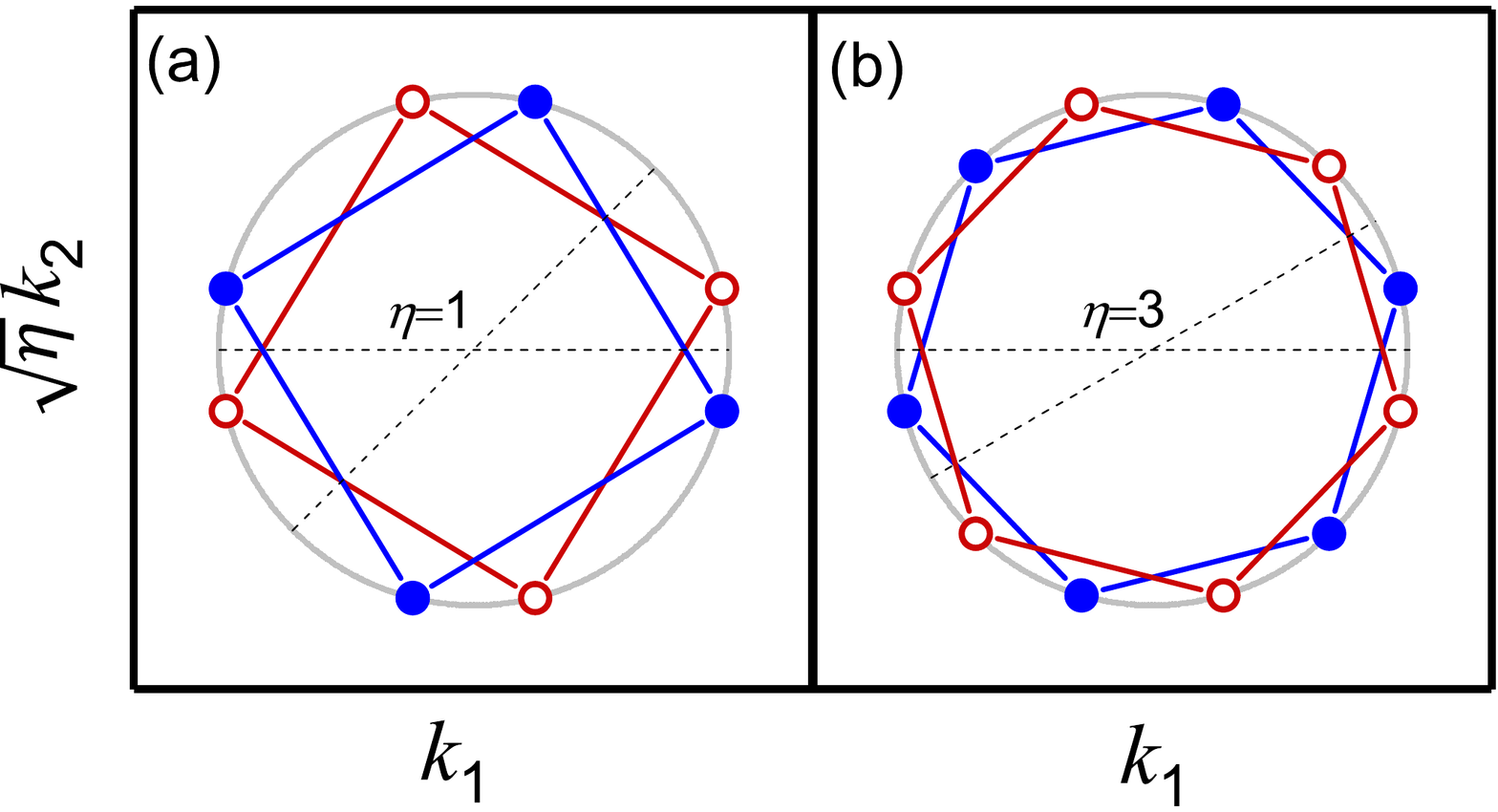}
\caption{Momentum Distributions of Classical Collision} 
{Momentum distributions in the phase space$\left(
k_{1},\protect\sqrt{\protect\eta }k_{2}\right) $ for two particles with mass
ratio (a) $\protect\eta =1$ and (b) $\protect\eta =3$, respectively. The
momentums are distributed on vertices of two polygons. While vertices on
each polygon fulfill $C_{4}$ and $C_{6}$ symmetries for (a) and (b),
respectively, vertices on different polygons can be transformed into each
other by axial reflection transformations. The dashed lines shown in (a) and
(b) are two reflection axes corresponding to axial reflection
transformations $\protect\sigma_z$ and $R \protect\sigma_z$, respectively.
(color online.)}
\label{fig2}
\end{figure}

\section*{SOLVABLE QUANTUM SYSTEM WITH IMBALANCED MASSES}

\section*{Model and Bethe-type-ansatz solution}

Consider a quantum system of two particles with masses $m_{1}$ and $m_{2}$
confined in a 1D hard-wall trap of length $L$. Two atoms interact with each
other via the potential $g\delta \left( x_{1}-x_{2}\right) $, where $\delta
\left( x\right) $\ is the Dirac delta function and $g$ is the interaction
strength. The Hamiltonian can be written as

\begin{equation}
H=-\frac{\hbar ^{2}}{2m_{1}}\frac{\partial ^{2}}{\partial x_{1}^{2}}-\frac{%
\hbar ^{2}}{2m_{2}}\frac{\partial ^{2}}{\partial x_{2}^{2}}+g\delta \left(
x_{1}-x_{2}\right) ,  \label{1}
\end{equation}%
and the wave function $\Psi \left( x_{1},x_{2}\right) $ satisfies the open
boundary condition
\begin{equation}
\Psi \left( x_{i}=\pm L/2\right) =0,  \label{boundry}
\end{equation}%
for $i=1$ and $2$. The system with equal mass reduces to the well-known
solvable Lieb-Liniger model under the open boundary condition \citep%
{Gaudin1971}. If the mass ratio fulfills the nonergodicity condition (\ref%
{nonergodicity}), the number of momentum vectors in the set is finite such
that the wavefunction of the quantum system can be taken in terms of Bethe-type
hypothesis as%
\begin{eqnarray}
\Psi (x_{1},x_{2}) &=&\theta \left( x_{2}<x_{1}\right)
\sum_{j}A_{j+}e^{i\left( d_{j}\mathbf{k}\right) ^{T}\mathbf{\cdot x}}  \notag
\\
&&+\theta \left( x_{1}<x_{2}\right) \sum_{j}A_{j-}e^{i\left( d_{j}\mathbf{k}%
\right) ^{T}\mathbf{\cdot x}},  \label{wfunction}
\end{eqnarray}%
where $A_{j\pm} $ are the coefficient of plane waves with different
quasimomentums and $\theta \left( x\right) $ is the step function. $\mathbf{x%
}=\left( x_{1},x_{2}\right) ^{T}$ and $\mathbf{k}=\left(k_{1},k_{2}\right)
^{T}$ are the coordinate vector and the quasimomentum vector of particle $1$
and $2$, respectively, and the collision operator $d_{j}\in D_{2n}$ with $%
j=1,2,\cdots ,4n$. Here $D_{2n}$ is the same dihedral group as in the classical
model in previous section. The wave function includes all possible terms in the scattering
process with the quasimomentums in the plane waves fulfilling the $D_{2n}$ symmetry. When $\eta=1$, we have $d_{j}\in D_{4}$ and Eq.(\ref{wfunction}) reduces to the Bethe ansatz wavefuntion of two-particle Lieb-Liniger model under the open boundary condition \citep{Gaudin1971}. Although the wavefunction of Eq.(\ref{wfunction}) is represented in a general form with $d_{j}\in D_{2n}$, in this work we only study the case with $d_{j}\in D_{6}$ corresponding to $\eta =3$ or $1/3$, as we find that it is the only exactly solvable example of quantum mass-imbalance systems with $\eta$ fulfilling the nonergodicity condition Eq.(\ref{nonergodicity}). In the following part, we focus on the $\eta =3$ case,
which occurs for example in a quantum gas with the formation of trimers with three times of the atomic mass.
The case of $\eta=1/3$ can be exactly solved within the same scheme due to the duality relation between $\eta$ and $1/\eta$. For other cases corresponding to irrational $\eta$, we can not find exact solutions by using the Bethe ansatz method.
The exact spectrum of the system is given for arbitrary interaction strength.

Firstly we consider the open boundary condition Eq. $\left( \ref{boundry}%
\right) $. The definition of reflection matrix for particle $1$ on the right
wall and on the left wall is
\begin{equation}
R_{j}\left( 1,\pm \right) =\frac{A_{j\pm }}{A_{\underline{j}\pm }}=-\exp %
\left[ \mp iL\sum_{l=1,2}d_{j}^{1l}k_{l}\right] ,  \label{op1}
\end{equation}%
where $\pm$ in $R$ respectively corresponds to the region $x_{2}<x_{1}$ or $%
x_{1}<x_{2}$, and the superscripts of $d$ indicate the matrix element in the
$\left(k_{1},k_{2}\right) ^{T}$ space. For convenience, we use $A_{j\pm }$
denotes the coefficients corresponding to the quasimomentum vector $%
(k_{1}^{\prime },k_{2}^{\prime }) = \left( d_{j}\mathbf{k}\right) ^{T}$,
where, for example, $k_{1}^{\prime }=\sum_{l=1,2}d_{j}^{1l}k_{l}$ denotes
the quasimomentum of particle $1$ after collision operator $d_{j}$. We
further let ${A_{\underline{j}\pm }}$ represent the coefficients
corresponding to the quasimomentum vector $(-k_{1}^{\prime },k_{2}^{\prime
}) = \left( d_{\underline{j}} \mathbf{k}\right) ^{T}$, where
\begin{equation}
d_{\underline{j}} = -\sigma _{z}d_{j},  \label{drel}
\end{equation}%
%
%For example, if $d_{\text{\b{j}}}=s\left( 3\right) \sigma _{z}=r$, then we can get $d_{j}=-\sigma _{z}s\left( 3\right) \sigma _{z}=r^{2}\sigma _{z}$, where we have used the property $\sigma _{z}r\sigma _{z}=r^{-1}$.
%For a given collision operator $d_{\text{\b{j}}}$, we always find another collision operator\ $d_{j}$ and make them satisfy relation $\left( \ref{drel}% \right) $.
and the underline of $j$ indicates the reflection of particle $1$. In a
similar way, we define the reflection matrix of particle $2$ on the left
wall and on the right wall as%
\begin{equation}
R_{j}\left( 2,\pm \right) =\frac{A_{j\pm }}{A_{\bar{j}\pm }}=-\exp \left[
\pm iL\sum_{l=1,2}d_{j}^{2l}k_{l}\right].  \label{op2}
\end{equation}%
%
%$k_{1}^{\prime }=\sum_{l=1,2}d_{j}^{1l}k_{l}$ is the momentum of particle $2$ after collision operator $d_{j}$ and the $j$th and $\bar{j}$th collision operator satisfy
Here $A_{\bar{j}\pm }$ represents the coefficients corresponding to the
quasimomentum vector $(k_{1}^{\prime },- k_{2}^{\prime }) = \left(d_{\bar{j}%
} \mathbf{k}\right) ^{T}$, where
\begin{equation}
d_{\bar{j}}= \sigma _{z}d_{j}  \label{dre2}
\end{equation}%
and the overline of $j$ indicates the reflection of particle $2$.

Next we discuss the scattering between two particles. In the relative
coordinate the first derivative of wave function is not continuous due to
the $\delta $ interaction. We integrate the Schr\"odinger equation with
Hamiltonian $\left( \ref{1}\right) $ from $x=-\varepsilon $ to $%
x=+\varepsilon $ and then take the limit $\varepsilon \rightarrow 0$. The
result is%
\begin{equation}
\left[ \frac{\partial \Psi }{\partial x}|_{x=0_{+}}-\frac{\partial \Psi }{%
\partial x}|_{x=0_{-}}\right] -\frac{2\mu }{\hbar ^{2}}g\Psi |_{x=0}=0 ,
\label{dis}
\end{equation}%
where the relative coordinate $x=x_{1}-x_{2}$ and the reduced mass $\mu
=m_{1}/\left( \eta +1\right) =m_{1}/4$. Inserting the Bethe-type-ansatz wave
function\ $\left( \ref{wfunction}\right) $ into Eq. $\left( \ref{dis}\right)
$, we get the relation
\begin{eqnarray}
&&i\left( \frac{d_{j}^{11}k_{1}+d_{j}^{12}k_{2}}{1+\eta }-\frac{%
d_{j}^{21}k_{1}+d_{j}^{22}k_{2}}{1+1/\eta }\right)  \notag \\
&&\times \left( A_{j+}-A_{k+}-A_{j-}+A_{k-}\right)  \notag \\
&=&\frac{2\mu }{\hbar ^{2}}g\left( A_{j-}+A_{k-}\right) ,  \label{ds1}
\end{eqnarray}%
where $A_{j \pm }$ and $A_{k \pm }$ represent the coefficients corresponding
to the quasimomentums $\left(d_{j} \mathbf{k}\right) ^{T}$ and $\left(d_{k}
\mathbf{k}\right) ^{T}$, respectively, and the collision operator $d_{j}$ is
related to $d_{k}$ via the relation
\begin{equation}
d_{j}=sd_{k}.  \label{resub}
\end{equation}%
On the other hand, for quasimomentum vector $(k_{1}^{\prime },k_{2}^{\prime
}) =\left(d_{j} \mathbf{k}\right) ^{T}$, there always exists $%
(-k_{1}^{\prime },-k_{2}^{\prime }) =\left(- d_{j} \mathbf{k}\right) ^{T}$.
We denote $d_{\breve{j}}=-d_{j}$ and $d_{\breve{k}}=-d_{k}$, and the
corresponding coefficients as $A_{\breve{j} \pm}$ and $A_{\breve{k} \pm}$,
which fulfills
\begin{eqnarray}
&&i\left( \frac{d_{\breve{j}}^{11}k_{1}+d_{\breve{j}}^{12}k_{2}}{1+\eta }-%
\frac{d_{\breve{j}}^{21}k_{1}+d_{\breve{j}}^{22}k_{2}}{1+1/\eta }\right)
\notag \\
&&\times \left( A_{\breve{j}+}-A_{\breve{k}+}-A_{\breve{j}-}+A_{\breve{k}%
-}\right)  \notag \\
&=&\frac{2\mu }{\hbar ^{2}}g\left( A_{\breve{j}-}+A_{\breve{k}-}\right) ,
\label{ds12}
\end{eqnarray}%
where $d_{\breve{j}}=sd_{\breve{k}}$ due to Eq. (\ref{resub}). Representing
\begin{equation}
A_{\breve{j} \pm}=T_{j,\pm }A_{j\pm},  \label{Tj}
\end{equation}
we have
\begin{eqnarray}
&&i\left( -\frac{d_{j}^{11}k_{1}+d_{j}^{12}k_{2}}{1+\eta }+\frac{%
d_{j}^{21}k_{1}+d_{j}^{22}k_{2}}{1+1/\eta }\right)  \notag \\
&&\times \left(
T_{j,+}A_{j+}-T_{k,+}A_{k+}-T_{j,-}A_{j-}+T_{k,-}A_{k-}\right)  \notag \\
&=&\frac{2\mu }{\hbar ^{2}}g\left( T_{j,-}A_{j-}+T_{k,-}A_{k-}\right) .
\label{ds22}
\end{eqnarray}%
From Eq.(\ref{Tj}), Eq.(\ref{op1}) and Eq. $\left( \ref{op2}\right) $, we
can get
\begin{equation}
T_{j,\pm }=\exp \left[ \mp iL \sum_{l=1,2} \left(
d_{j}^{2l}k_{l}-d_{j}^{1l}k_{l}\right) \right],
\end{equation}%
such that the Eq. $\left( \ref{ds22}\right) $ actually represents the second
relation between $A_{j\pm }$ and $A_{k\pm }$.

The continuity of the wavefunction $\Psi |_{x=0_{+}}=\Psi |_{x=0_{-}}$ gives
yet another pair of equations
\begin{equation}
A_{j+}+A_{k+}=A_{j-}+A_{k-}  \label{c1}
\end{equation}%
and
\begin{equation}
T_{j,+}A_{j+}+T_{k,+}A_{k+}=T_{j,-}A_{j-}+T_{k,-}A_{k-}.  \label{c12}
\end{equation}

Obviously the two-particle scattering problem with unequal mass is much more
complicated than the equal mass case. For each pair of $k$ and $j$ related
by Eq.(\ref{resub}), we have four homogeneous linear equations of four
coefficients $A_{j\pm }$ and $A_{k\pm }$, given by Eq. $\left( \ref{ds1}%
\right) $, $\left( \ref{ds22}\right) $, $\left( \ref{c1}\right) $, and $%
\left( \ref{c12}\right) $. Non-trivial solution of these coefficients
requires the determinant of the corresponding matrix equations to be zero.
%The 24 elements in group $D_{6}$ thus give us 6 equations, 
Since $d_j \in D_{6}$, there are altogether $24$ different coefficients and we can get $6$ equations,
among which three
equations are identical to the other three. This leads to the constraint of the
momentum $k_{1}$ and $k_{2}$, which can be shown to be equivalent to either
the following pair of Bethe-type-ansatz equations
\begin{equation}
\left\{
\begin{array}{c}
k_{1}+3k_{2}-\frac{2\mu }{\hbar ^{2}}g\left( \cot \frac{\left(
k_{1}+k_{2}\right) L}{2}+\cot k_{2}L\right) =0, \\
k_{1}-3k_{2}-\frac{2\mu }{\hbar ^{2}}g\left( \cot \frac{\left(
k_{1}-k_{2}\right) L}{2}-\cot k_{2}L\right) =0,%
\end{array}%
\right.  \label{BA1}
\end{equation}%
or that of
\begin{equation}
\left\{
\begin{array}{c}
k_{1}+3k_{2}+\frac{2\mu }{\hbar ^{2}}g\left( \tan \frac{\left(
k_{1}+k_{2}\right) L}{2}+\tan k_{2}L\right) =0, \\
k_{1}-3k_{2}+\frac{2\mu }{\hbar ^{2}}g\left( \tan \frac{\left(
k_{1}-k_{2}\right) L}{2}-\tan k_{2}L\right) =0.%
\end{array}%
\right.  \label{BA2}
\end{equation}%

\begin{figure}[tbp]
\includegraphics[width=0.75\textwidth]{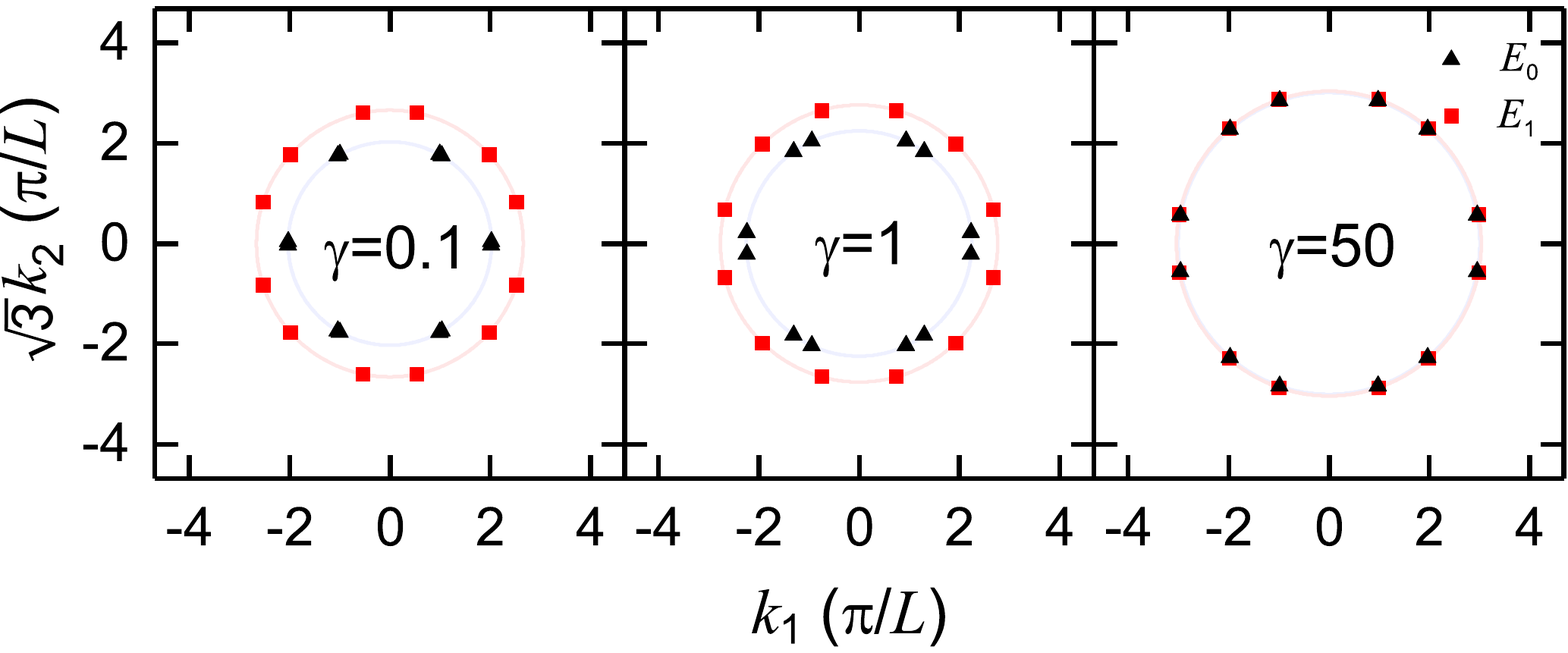}
\caption{Quasimomentums $k_{1}$ and $k_{2}$ of Quantum System} 
{Quasimomentums $k_{1}$ and $k_{2}$ for the ground
state and the first excited state with mass ratio $\protect\eta =3$ for
different interaction strengths $\protect\gamma =0.1$, $1$ and $50$. The
triangles and squares show the quasimomentums of the ground state and the
first excited state, respectively. (color online.)}
\label{fig3}
\end{figure}

Here we would like to add a remark for the other mass-imbalanced cases, for example, the case with $\eta=3-2\sqrt{2}$. In this case, we have $d_j \in D_8$ and there are altogether $32$ different coefficients. By setting the determinant of corresponding matrix equations of Eqs.(24), (28), (30), and (31) to zero, we can get $8$ equations  with four of them being identical to the other four. So there are four independent equations with two undetermined variables $k_{1}$ and $k_{2}$, which generally yields no solutions, i.e., it is impossible for $k_{1}$ and $k_{2}$ to fulfill four independent equations simultaneously. This means that the Bethe-type-ansatz wavefunction given by Eq. (\ref{wfunction}) is not the eigenstate of the mass-imbalanced system with $\eta=3-2\sqrt{2}$.  We also verified analytically that for this mass ratio there is no solution in the hard-core interacting case $g \rightarrow +\infty$.

\section*{Results and discussions}

By numerically solving the transcendental equations $\left( \ref{BA1}\right)
$ or $\left( \ref{BA2}\right) $, we can get the quasimomentums for any given
interaction $g$. Contrary to the classical model where the momentum can take
any continuous values, the quasimomentum here is quantized and can only take
discrete values. For convenience, we introduce the dimensionless interaction
strength $\gamma =\frac{\mu gL}{\hbar ^{2}}$ and adopt the natural units $%
\hbar =\mu =L=1$. In Fig. 3, we display the solution of
quasimomentums for different $\gamma $ when the system is in the ground
state and the first excited state. For a finite $\gamma $, the
quasimomentums in the figure can be generally classified into three groups,
denoted as $(\pm k_{1},\pm k_{2})$, $(\pm k_{1}^{\prime },\pm k_{2}^{\prime
})$, and $(\pm k_{1}^{\prime \prime },\pm k_{2}^{\prime \prime })$,
respectively. The phase space can be divided into four quadrants and every
quadrant is sprinkled with three points. The points in the first quadrant
are related to those in other quadrants by the reflection operators $\sigma
_{z}$, $-\sigma _{z}$ and $-I$. Thus we focus on the three points in first
quadrant: $\mathbf{k}=(k_{1},k_{2})^{T}$, $\mathbf{k}^{\prime
}=(k_{1}^{\prime },k_{2}^{\prime})^T$ and $\mathbf{k}^{\prime \prime
}=(k_{1}^{\prime \prime },k_{2}^{\prime \prime})^T$. Once we find a solution
$\mathbf{k}$ from the Bethe-type-ansatz equations, it is easy to obtain $\mathbf{k%
}^{\prime }$ and $\mathbf{k}^{\prime \prime }$ by applying appropraite group
operators on $\mathbf{k}$, which necessarily fulfill the same equations.
Take the ground state for $\gamma =1$ in Fig. 3(b) as an example.
From $\mathbf{k}=(0.93667\pi ,1.17904\pi )^{T}$, one immediately knows that $%
\mathbf{k}^{\prime }=-\sigma _{z}s\sigma _{z}\mathbf{k}=(1.30023\pi
,1.05786\pi )^{T}$ and $\mathbf{k}^{\prime \prime }=\sigma _{z}s\mathbf{k}%
=(2.2369\pi ,0.12119\pi )^{T}$ are all the solution of Bethe-type-ansatz
equations $\left( \ref{BA1}\right) $.

Particularly, when $\gamma \rightarrow 0$, we find every two points of
momentum in the ground state tend to be the same and $1/3$ of the points
will be located on the straight line $k_2=0$. Note that for the
non-interacting case, the transcendental equations $\left( \ref{BA1}\right) $
reduces to
\begin{equation}
\left\{
\begin{array}{c}
\cot \frac{ k_{1}+k_{2}}{2}+\cot k_{2} = \infty, \\
\cot \frac{ k_{1}-k_{2}}{2}-\cot k_{2} = \infty,%
\end{array}%
\right.
\end{equation}%
%
%
%
%
%
%
%
%
%
%
%or equivalently
%\begin{eqnarray}
% \frac{\sin  \frac{\left(
%3 k_{1}+k_{2}\right) L}{2}}{\sin k_2L \sin  \frac{\left(
% k_{2}+k_{1}\right) L}{2}} &=& \infty, \\
%\frac{\sin  \frac{\left(
%3 k_{2}-k_{1}\right) L}{2}}{\sin k_2L \sin  \frac{\left(
% k_{2}-k_{1}\right) L}{2}}  &=& \infty,
%\end{eqnarray}%
which leads to the single particle solution: $k_1=n_1\pi$, $k_2=n_2\pi$,
where $n_1$ and $n_2$ are integers. The quantum numbers of the ground state
is $(n_{1},n_{2})=(1,1)$ and the corresponding energy is $\pi^2/2$.
%The quantum numbers of the ground
%state and lowest few excited states are $(n_{1},n_{2})=(1,1)$, $(2,1)$, $(3,1)$, $(1,2)$, etc.
%with energies in unit of $\pi^2/2m_1L^2$ are $4, 7, 12, 13,$ etc.
We find that $\mathbf{k}=\mathbf{k}^{\prime}=(\pi,\pi)^T$ are equal, which
holds for all other cases with $n_1=n_2$. The coefficients for the plane
waves with $\mathbf{k}^{\prime \prime}=(2\pi,0)^T$, however, are vanishing
in the non-interacting case and the wave function is but the direct product
state of the two single-particle ground states. The plane waves with
approximately $(\pm 2\pi, 0)$, i.e. the four points near the $k_2$ axis,
prove to be emergent solutions uniquely in the weakly interacting case, as
the wave function of zero momentum, that is, a constant, violates the
vanishing condition at both left and right boundaries in the non-interacting
case. More emergent solutions like these are found for the excited states,
which are prohibited in the non-interacting case and yet contribute in the
superposition of Bethe-type hypothesis of the interacting many-body wave
function. For instance, the emergent solutions for the first excited state
corresponding to $(n_{1},n_{2})=(2,1)$ with approximate energy $7\pi^2/8$
are plane waves with the momentum taking the values near
half-integer-multiple of $\pi$, specifically, $\mathbf{k}^{\prime} \approx
(5\pi/2,\pi/2)^T$, $\mathbf{k}^{\prime \prime } \approx (\pi/2,3\pi/2)^T$.
This is an intrinsic feature for the mass-imbalanced system as we have
noticed that no solutions emerge in the equal mass case.

On the other hand, when $\gamma \rightarrow \infty$, the ground state and
the first excited state tend to be degenerate. In this case the
transcendental equations $\left( \ref{BA1}\right)$ reduce to
\begin{equation}
\left\{
\begin{array}{c}
\cot \frac{ k_{1}+k_{2}}{2}+\cot k_{2} = 0, \\
\cot \frac{ k_{1}-k_{2} }{2}-\cot k_{2} = 0.%
\end{array}%
\right.  \label{infinite}
\end{equation}%
From the above equations, it follows that $k_1= n_1 \pi$ and $k_2= n_2 \pi/3$
with $n_1$ and $n_2$ being integers. The symmetry in the quasi-momentum set,
however, constraints the values of $n_1$ and $n_2$ to some specific
integers. This can be understood as following: in the momentum set, not only
$\mathbf{k}$, but also $\mathbf{k}^{\prime }$ and $\mathbf{k}^{\prime \prime
}$, which are related by collision operators in the group $D_6$, necessarily
satisfy the above equations. For example, when $n_1=1$, $n_2=1$, the
momentum values $\mathbf{k}^{\prime }=s\sigma _{z}\mathbf{k}=(0, 2\pi/3)^T$
violate the equations (\ref{infinite}), while $n_1=1$, $n_2=2$, $\mathbf{k}%
^{\prime }=s\mathbf{k}=(3\pi/2, \pi/6)^T$ again fail them, etc. It can be
shown when $n_1=1$, the minimum value of $n_2$ to satisfy (\ref{infinite})
is $n_2=5$. So the lowest values for the quasimomentum are $\mathbf{k}=(\pi,
5\pi/3)^T$, $\mathbf{k}^{\prime}=(2\pi, 4\pi/3) ^ T$ and $\mathbf{k}^{\prime
\prime} = (3\pi, \pi/3)^ T$. In the infinitely interacting case, the ground
state and the first excited state are degenerate with eigenenergy $7\pi^2/6$
which is a little bit larger than the first excited state energy of the
non-interacting case.
% and second and third excited states are  $(k_1, k_2)=(1, 7/3)\pi/L$, $(k_1', k_2')=(3, 5/3)\pi/L$, $(k_1'', k_2'')=(4, 2/3)\pi/L$, respectively.
The solutions at these two limits are consistent with the alternative
analysis in the transparent method  B.

\begin{figure}[tbp]
\includegraphics[width=0.75\textwidth]{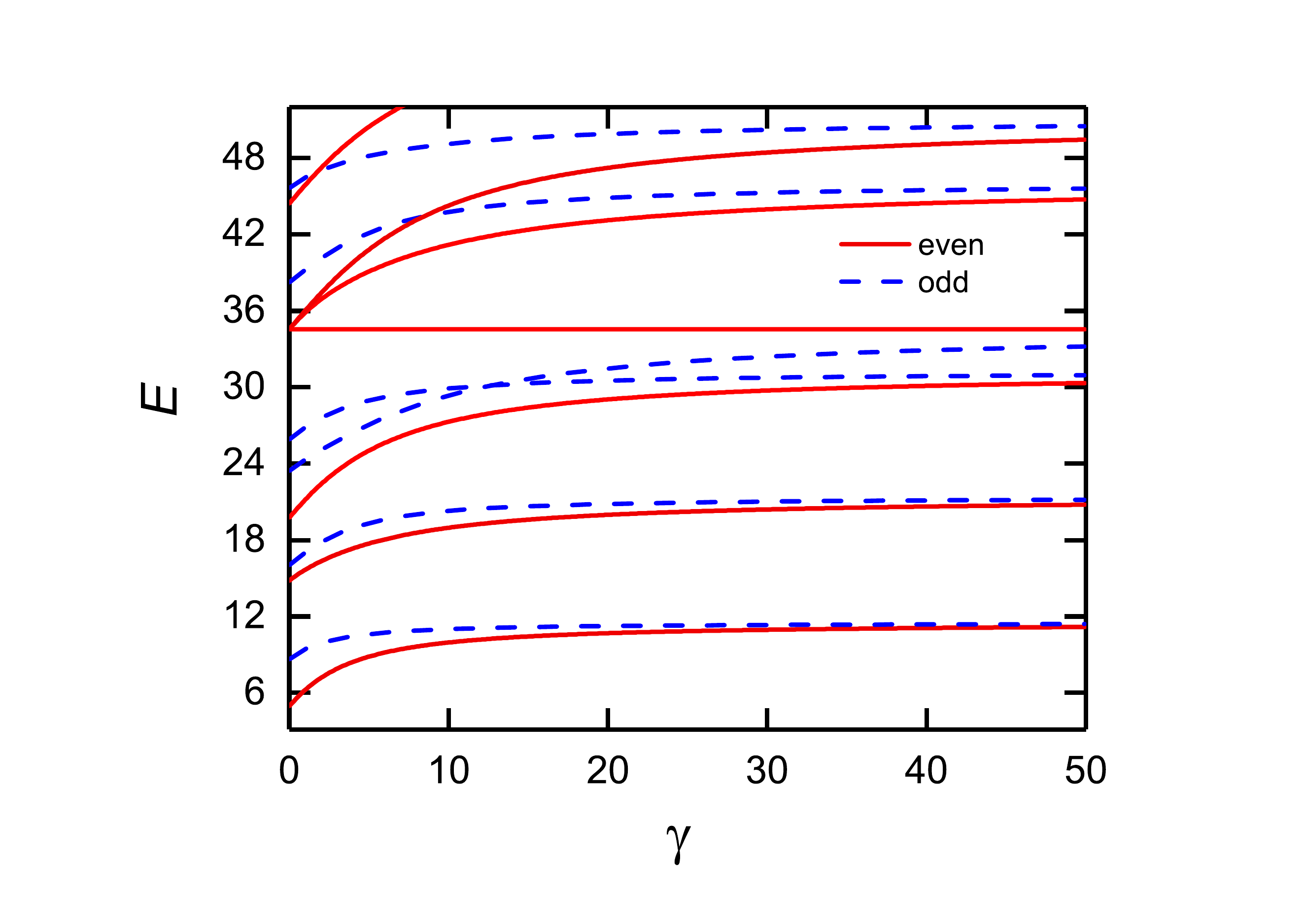}
\caption{Energy Spectrum of Two Atoms with Mass Ratio $%
\protect\eta =3$ in the Hard-wall Trap as a Function of the Interaction
Strength $\protect\gamma $}
{The red solid lines show the eigenstates with
even parity and the blue dashed lines with odd parity. (color online.)}
\label{fig4}
\end{figure}

The finite interaction case interpolates between these two limits as shown
in Fig. $3$. We find that the momentum points in the weak
interaction case $\gamma =0.1$ are very close to the free particle case $%
\gamma =0$. Nevertheless, the heavy and light particles in the interacting
case are entangled and the wave function is no longer a product state. The
quasimomentum points for the ground state occupy the vertices of two regular
hexagons on a circle in the phase space $(k_1,\sqrt{3}k_2)$, while the
overlapped points in the free particle case start to be split into two when
the interaction gradually sets in. The ground state circle then expands
towards that of the first excited state with the increase of the interaction
strength and finally join it in the infinitely interacting case, leading to
the degeneracy of the two states, which is already clearly seen for $%
\gamma=50$ as shown in Fig. $3$(c).

In Fig. $4$, we plot the energy spectrum $E=(k_{1}^{2}+3k_{2}^{2})/8$
as a function of the interaction $\gamma $. We find that every energy level
corresponds to a fixed parity, as the corresponding wavefunction fulfills
the parity symmetry:
\begin{equation}
\Psi(x_{1},x_{2})=\pm \Psi (-x_{1},-x_{2}),
\end{equation}
where the even parity is with sign "$+$" and odd parity with "$-$". With the
increase of $\gamma$, the eigenvalues generally increase except for some
special states, e.g., the seventh excited state as shown in Fig. 5
does not change with $\gamma$. In the limit case $\gamma =\infty $, two
levels with opposite parity tend to be doubly degenerate, and the wave
functions vanish along the line $x_1=x_2$.

We note that the seventh excited state is an even parity state whose energy
is independent of the interaction strength. The existence of such a state is
related to the emergence of a triple degenerate point in the noninteracting
limit $\gamma =0$. These three degenerate states are labeled by quantum
numbers $(n_{1},n_{2})=(5,1)$, $(4,2)$, and $(1,3)$, respectively, which
have no correspondence in the equal mass case. In the presence of
interaction, the triple degeneracy is usually broken. Nevertheless, we can
construct a wavefunction composed of a superposition of triple degenerate
eigenstates, which is the eigenstate of the interacting Hamiltonian with
eigenvalue irrelevant to the interaction strength. Explicitly, the
wavefunction of this state is given by
\begin{eqnarray}
\Psi (x_{1},x_{2}) &=&\frac{1}{\sqrt{3}}[\phi _{5}(x_{1})\phi _{1}(x_{2})
\notag \\
&&-\phi _{4}(x_{1})\phi _{2}(x_{2})+\phi _{1}(x_{1})\phi _{3}(x_{2})],
\label{8th-state}
\end{eqnarray}%
where
\begin{equation*}
\phi _{n}\left( x\right) =\sqrt{\frac{2}{L}}\sin \frac{n\pi }{L}\left( \frac{%
L}{2}+x\right)
\end{equation*}%
is the $n$-th single-particle eigenstate of the hard well.
%The answer goes back to the equation $\left( \ref{dis}\right)$, which is the only condition concerning the interaction strength.
After some straightforward algebras, it is easy to check that $\Psi
(x_{1},x_{2})|_{x_1=x_2}=0$, i.e., the wavefunction takes zero at $%
x_{1}=x_{2}$, indicating that the state given by Eq. (\ref{8th-state}) is
the eigenstate of Hamiltonian (\ref{1}) irrelevant to the value of $\gamma$.
Actually, there are a series of such excited states, corresponding to the
higher triple degenerate points in the noninteracting limit. Generally,
triple degenerate states are characterized by the quantum numbers $%
(n_{1},n_{2})$, which should fulfill three conditions, i.e., $n_{1}+n_{2}$
is even, $n_{1}\neq n_{2}$ and $n_{1}\neq 3n_{2}$. The corresponding
wavefunction can be written as%
\begin{eqnarray*}
\Psi (x_{1},x_{2}) &=&\frac{1}{\sqrt{3}}[\phi _{n_{1}}(x_{1})\phi
_{n_{2}}(x_{2}) \\
&&\pm \phi _{n_{1}^{\prime }}(x_{1})\phi _{n_{2}^{\prime }}(x_{2})\pm \phi
_{n_{1}^{\prime \prime }}(x_{1})\phi _{n_{2}^{\prime \prime }}(x_{2})],
\end{eqnarray*}%
where the selections of $"\pm "$ depend on the concrete values of quantum
numbers $n_{1}$ and $n_{2}$. An example for the next excited state,
independent of $\gamma$, is labeled by quantum numbers $(n_{1},n_{2})=(2,4)$%
, $(n_{1}^{\prime },n_{2}^{\prime })=(7,1)$, and $(n_{1}^{\prime \prime
},n_{2}^{\prime \prime})=(5,3)$.

%With the exact solution at hand, we can explore many correlation and dynamical properties of this system. For instance, the two-body density matrix
In Fig. $5$, we display the probability density distribution $\rho
\left( x_{1},x_{2}\right) =\left\vert \Psi (x_{1},x_{2})\right\vert ^{2}$
for the ground state and the first excited state as well as the seventh
excited state with three typical interaction strength parameters $\gamma
=0.1, 1, 10$.
%The node of the wave function along $x_1$, that is, the heavy particle, increases from 0 to 2 in the lowest 3 levels, while the first node along $x_2$, for the light particle, occurs only in the 4th level.
Comparing to the equal mass case, the two-body wavefunction no longer has
exchange symmetry, nevertheless it keeps the parity symmetry. It is obvious
that the density distribution fulfills $\rho \left( -x_{1},-
x_{2}\right)=\rho \left( x_{1},x_{2}\right)$. We find that with the
interaction increased, particles will avoid to occupy the same position and
the density along the diagonal line $x_{1}=x_{2}$ is greatly suppressed. In
the strong interaction region, the density of the ground and first excited
states exhibit almost the same density patterns. In the infinitely repulsive
limit, the densities for the degenerate states are exactly same with zero
distribution along the diagonal line. For the seventh excited state, it is
clear that the density distribution is independent of $\gamma$ and always
gives zero along the diagonal line.

\begin{figure}[tbp]
\includegraphics[width=0.75\textwidth]{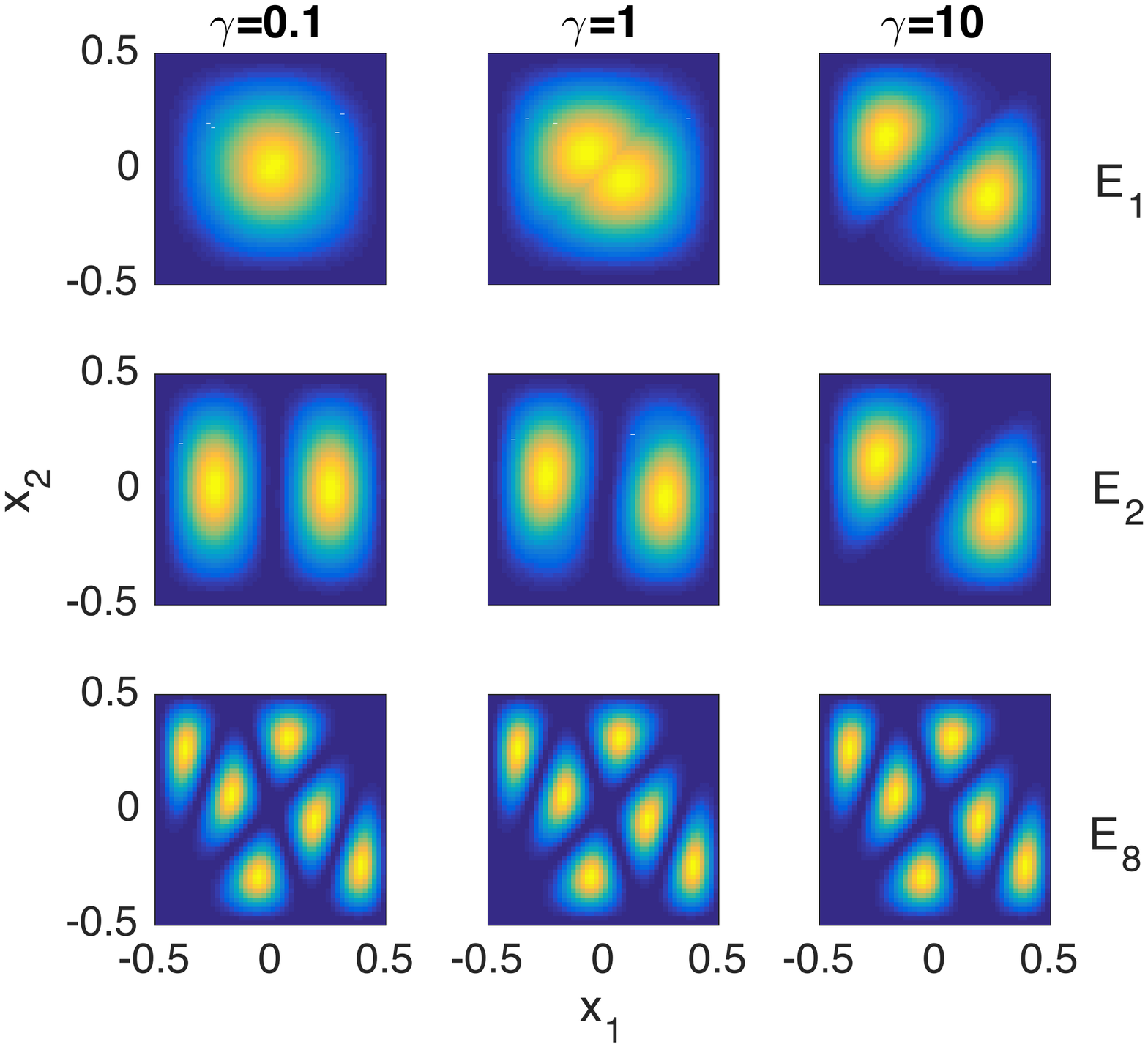}
\caption{The Normalized Probability Density $\protect\rho%
\left(x_1,x_2\right) $ for Two Unequal Mass Particles in the Hard-wall Trap}
{The columns represent results for three interaction parameters $\protect%
\gamma=0.1$, $1$ and $10 $, respectively. The rows from top to bottom are
for the ground state, the first and the $7$-th excited state, respectively. (color online.)}

\label{fig5}
\end{figure}

In summary, we study the problem of two interacting particles with unequal
masses in a hard-wall trap and unveil that the system is exactly solvable by
using Bethe-type ansatz only for the mass ratio $\eta=3$ or $1/3$. Since the
Bethe-type ansatz is based on the wavefunction hypothesis which requires finite
superpositions of plane waves, the solvability of the mass-imbalance quantum
system is thus related to a problem of seeking nonergodicity conditions in
the classical elastic collision in a 1D hard-wall trap. In general, each
collision and reflection process of two particles with unequal masses gives
rise to a new set of momentums $k_1$ and $k_2$, which shall not form finite
momentum distributions after multiple collisions. Nevertheless, we find that
finite momentum distributions after multiple collisions are available at
specific values of mass ratio, which is determined by the nonergodicity
condition. For $\eta=3$ or $1/3$, the
permitted momentums fulfill the $D_{6}$ symmetry. Based on the Bethe-type ansatz,
we then exactly solve the quantum system with mass ratio $3$ and
give Bethe-type-ansatz equations for arbitrary
interaction strength. By solving the Bethe-type-ansatz equations, we give the
energy spectrum and wavefunctions of the mass-imbalance system with $\eta=3$, which are
found to display some peculiar behaviors with no correspondence in the
equal-mass system.

\section*{Limitation of the Study}

Although nonergodicity condition for the classical collision problem in a hard-wall trap includes a
series of solutions of mass ratio, the extended Bethe ansatz method can only give the exact solution
for the two-particle quantum system with the mass ratio $\eta = 3$ or $1/3$. Our method can not be directly 
applied to solve the three-particle system. The properties of many-particle interacting models with
unequal masses are still not clear and worth further investigating.

\section*{Methods}
All methods can be found in the accompanying {\bf {Transparent Methods supplemental file}}.

\section*{Supplemental information}
Supplemental Information includes Transparent Methods.

\section*{Acknowledgments}
We thank Y. C. Yu and B. Wu for helpful discussions. This work is supported
by NSFC under Grants No. 11674201 and the National Key
Research and Development Program of China (2016YFA0300600 and
2016YFA0302104).

\section*{Author contributions}
 Y.Z. and S. C. conceived the project and supervised the study. Y.L. and F.Q. performed the numerical 
 calculations and theoretical analysis. Y.L., Y.Z., and S. C. contributed to writing the manuscript.

\section*{Declaration of interests} 
The authors declare no competing interests.

\section*{Supplemental Information}

\appendix

\renewcommand{\theequation}{S\arabic{equation}}

\section*{Transparent Methods}

\section*{A. Solution to equations for nonergodicity condition}

Here we show the derivation of the nonergodicity condition 
\begin{equation}
\eta =\tan ^{2}l\pi /2n,  \label{Snonergodicity}
\end{equation}%

 by solving the equations
 
\begin{equation}
\left( s\left( \eta \right) \sigma _{z}\right) ^{n}\left(
\begin{array}{c}
k_{1} \\
k_{2}%
\end{array}%
\right) =\pm \left(
\begin{array}{c}
k_{1} \\
k_{2}%
\end{array}%
\right)  \label{SC1}
\end{equation}%
and
\begin{equation}
\left( \sigma _{z} s\left( \eta \right)\right) ^{n}\left(
\begin{array}{c}
k_{1} \\
k_{2}%
\end{array}%
\right) =\pm \left(
\begin{array}{c}
k_{1} \\
k_{2}%
\end{array}%
\right). \label{SC2}
\end{equation}%

%\begin{equation}
%\left( s\left( \eta \right) \sigma _{z}\right) ^{n}\left(
%\begin{array}{c}
%k_{1} \\
%k_{2}%
%\end{array}%
%\right) =\pm \left(
%\begin{array}{c}
%k_{1} \\
%k_{2}%
%\end{array}%
%\right)  \label{eq}
%\end{equation}%
%or%
%\begin{equation*}
%\left( \sigma _{z}s\left( \eta \right) \right) ^{n}\left(
%\begin{array}{c}
%k_{1} \\
%k_{2}%
%\end{array}%
%\right) =\pm \left(
%\begin{array}{c}
%k_{1} \\
%k_{2}%
%\end{array}%
%\right) .
%\end{equation*}%
A quite useful tool, Chebyshev identity, is used to derive the relation for
the matrix elements of the $n$th power of the matrix.

Consider a unimodular matrix $\mathbf{M}$ given by
\begin{equation}
\mathbf{M}=\left(
\begin{array}{cc}
a & b \\
c & d%
\end{array}%
\right) ,
\end{equation}%
where $\text{Det}\mathbf{M}=ad-bc=1$. Suppose that eigenvalues of the
unimodular matrix $\mathbf{M}$ are given by
\begin{equation}
\lambda _{1}=e^{iq}\text{ and }\lambda _{2}=e^{-iq},\text{ }
\end{equation}%
then the $n$-th power of the matrix $\mathbf{M}$ can be represented as \citep%
{Yeh}
\begin{equation}
\mathbf{M}^{n}=\left(
\begin{array}{cc}
a & b \\
c & d%
\end{array}%
\right) ^{n} \\
=\left(
\begin{array}{cc}
aU_{n-1}-U_{n-2} & bU_{n-1} \\
cU_{n-1} & dU_{n-1}-U_{n-2}%
\end{array}%
\right) ,  \label{cheb}
\end{equation}%
where the function $U_{n}$ is defined as%
\begin{equation}
U_{n}=\frac{\sin \left( n+1\right) q}{\sin q},  \label{sol}
\end{equation}%
and $q$ is given by the eigenvalues of the matrix $\mathbf{M}$ via the
relation
\begin{equation}
Tr\mathbf{M=}\lambda _{1}+\lambda _{2}=2\cos q.  \label{tri}
\end{equation}%
The details for the derivation of the Chebyshev identity Eq. (\ref{cheb})
can be found in Ref. \citep{Yeh}.

Now we let%
\begin{equation}
\mathbf{M=}s\left( \eta \right) \sigma _{z}=\left(
\begin{array}{cc}
\frac{\eta -1}{\eta +1} & \frac{-2\eta }{\eta +1} \\
\frac{2}{\eta +1} & \frac{\eta -1}{\eta +1}%
\end{array}%
\right)
\end{equation}%
and%
\begin{equation}
\mathbf{M}^{\prime }\mathbf{=}\sigma _{z}s\left( \eta \right) =\left(
\begin{array}{cc}
\frac{\eta -1}{\eta +1} & \frac{2\eta }{\eta +1} \\
\frac{-2}{\eta +1} & \frac{\eta -1}{\eta +1}%
\end{array}%
\right) .
\end{equation}%
It is easy to check $\text{Det} \mathbf{M}=\text{Det} \mathbf{M}^{\prime }=1$%
. Comparing $\mathbf{M}$ and $\mathbf{M}^{\prime }$, we find that\ the
diagonal terms are the same, i.e. $\mathbf{M}_{11}=\mathbf{M}_{11}^{\prime }$%
, $\mathbf{M}_{22}=\mathbf{M}_{22}^{\prime }$ and off-diagonal terms are
opposite numbers with each other, i.e. $\mathbf{M}_{12}=-\mathbf{M}%
_{12}^{\prime }$, $\mathbf{M}_{21}=-\mathbf{M}_{21}^{\prime }$. So the
eigenvalues for two matrices are the same and can be represented as
\begin{equation}
\lambda _{1,2}=\frac{\eta -1\pm 2\sqrt{-\eta }}{\eta +1}.
\end{equation}%
From (\ref{tri}), we can get the relation
\begin{equation}
\cos q=\frac{\eta -1}{\eta +1}.  \label{cose}
\end{equation}%
To solve the equation (\ref{SC1}) or (\ref{SC2}) is equivalent to solve%
\begin{equation}
\left( \mathbf{M}\right) ^{n}=\pm \left(
\begin{array}{cc}
1 & 0 \\
0 & 1%
\end{array}%
\right)
\end{equation}%
or%
\begin{equation}
\left( \mathbf{M}^{\prime }\right) ^{n}=\pm \left(
\begin{array}{cc}
1 & 0 \\
0 & 1%
\end{array}%
\right).
\end{equation}%
Using the Chebyshev identity (\ref{cheb}), we can find that $\left( \mathbf{M%
}^{n}\right) _{11}=\left( \mathbf{M}^{n}\right) _{22}=\left( \mathbf{M}%
^{\prime n}\right) _{11}=\left( \mathbf{M}^{\prime n}\right) _{22}$. The
solutions of $\mathbf{M}$ and $\mathbf{M}^{\prime }$ satisfy the same
relation
\begin{equation*}
\left( \mathbf{M}^{n}\right) _{11}=\pm 1,
\end{equation*}%
this is%
\begin{equation}
\frac{\eta -1}{\eta +1}U_{n-1}-U_{n-2}=\cos nq=\pm 1.  \label{m11}
\end{equation}%
The solutions of (\ref{m11}) are
\begin{equation*}
q=\frac{l\pi }{n},l=1,2,3\cdots .
\end{equation*}%
Then we can also get%
\begin{equation*}
U_{n-1}=\frac{\sin nq}{\sin q}=0,
\end{equation*}%
which ensures that the off-diagonal terms of $\mathbf{M}$ and $\mathbf{M}%
^{\prime }$ are $0$. Solving the equation (\ref{cose}), we get
\begin{equation}
\eta =\frac{1+\cos q}{1-\cos q}=\frac{1}{\tan ^{2}\frac{l\pi }{2n}}%
,l=1,2,3\cdots .  \label{solu}
\end{equation}%
Because $l$ and $n$ are both integers, (\ref{solu}) can be written as other
form
\begin{equation*}
\eta =\frac{1}{\tan ^{2}\frac{\left( n-l\right) \pi }{2n}}=\tan ^{2}\frac{%
l\pi }{2n},l=1,2,3\cdots .
\end{equation*}%
The solutions requires that the diagonal terms of matrix $\mathbf{M}^{n}$($%
\mathbf{M}^{\prime n}$) equal $\pm 1$ and off-diagonal terms equal $0$, so
the sign of the off-diagonal do not affect the solutions.

\section*{B. The hard-core limit and $g=0$ limit}

We consider two limit cases. The first case is the hard-core limit with $%
g=\infty$, in which the wave function satisfies the boundary condition
\begin{equation*}
\Psi |_{x_{1}=x_{2}}=0.
\end{equation*}%
Inserting the Bethe-type wave function into the
above equation, we get a pair of equations:
\begin{equation}
A_{j\pm }=-A_{k\pm },d_{j}=sd_{k}  \label{coeA1}
\end{equation}%
and
\begin{equation}
T_{j,\pm }A_{j\pm }=-T_{k,\pm }A_{k\pm }.  \label{coeA2}
\end{equation}%
Combining Eq. (\ref{coeA1}) with Eq. (\ref{coeA2}), we get the relation
\begin{equation*}
\frac{T_{j,-}}{T_{k,-}}=1,
\end{equation*}%
i.e.,%
\begin{equation*}
\exp \left(
iL\sum_{l=1,2}3d_{j}^{2l}k_{l}-iL\sum_{l=1,2}d_{j}^{1l}k_{l}\right) =1,
\end{equation*}%
which gives rise to three independent equations:%
\begin{eqnarray*}
\exp \left[ iL\left( 3k_{2}-k_{1}\right) \right] &=& 1, \\
\exp \left[ iL\left( 3k_{2}+k_{1}\right) \right] &=& 1, \\
\exp \left[ 2iLk_{1}\right] &=& 1.
\end{eqnarray*}%
By solving the above equations, we can get a series of solution $%
k_{1}=l_{1}\pi /L$ and $k_{2}=l_{2}\pi /3L$, where $l_{1}$ and $l_{2}$ are
integers and some of them are redundant. Given that $\mathbf{k}=\left(
k_{1},k_{2}\right) ^{T} $ is a solution of the above transcendental
equations, all the quasimomentums obtained via $d_{j}\mathbf{k}$ should also
be the solution of transcendental equations, which gives some restrictions
to the values of $l_{1}$ and $l_{2}$. According to the ratio relations of
the coefficients described by reflection matrixes and Eq. (\ref{coeA1}), 
the wavefunction can be written as%
\begin{eqnarray*}
\Psi (x_{1},x_{2}) &=&\theta \left( x_{2}<x_{1}\right) [\Phi _{\mathbf{k}%
}(x_{1},x_{2}) \\
&&-\Phi _{\mathbf{k}^{\prime }}(x_{1},x_{2})+e^{-ik_{2}L}\Phi _{\mathbf{k}%
^{\prime \prime }}(x_{1},x_{2})] \\
&&\pm \theta \left( x_{1}<x_{2}\right) [\Phi _{\mathbf{k}}(-x_{1},-x_{2}) \\
&&-\Phi _{\mathbf{k}^{\prime }}(-x_{1},-x_{2})+e^{-ik_{2}L}\Phi _{\mathbf{k}%
^{\prime \prime }}(-x_{1},-x_{2})]
\end{eqnarray*}%
where $\mathbf{k}^{\prime }=s\mathbf{k}$, $\mathbf{k}^{\prime \prime
}=s\sigma_{z}\mathbf{k}$ and
\begin{eqnarray}
&&\Phi _{\mathbf{k}}(x_{1},x_{2})  \notag \\
&=&e^{i\left( k_{1}x_{1}+k_{2}x_{2}\right) }-e^{-ik_{2}L}e^{i\left(
k_{1}x_{1}-k_{2}x_{2}\right) }  \notag \\
&&-e^{ik_{1}L}e^{i\left( -k_{1}x_{1}+k_{2}x_{2}\right)
}+e^{-ik_{2}L+ik_{1}L}e^{i\left( -k_{1}x_{1}+k_{2}x_{2}\right) }  \notag \\
&=&4e^{-i\frac{k_{1}+k_{2}}{2}L}\sin k_{1}\left( \frac{L}{2}-x_{1}\right)
\sin k_{2}\left( \frac{L}{2}+x_{2}\right) .  \label{reform}
\end{eqnarray}
It is interesting to note that two mass-imbalance hard-core particles moving
in a 1D box is equivalent to a triangle billiard system \citep%
{WuBiao,WangJiao,Gorin}, and thus our exact result is helpful for
understanding quantum billiard systems from a different perspective.

The other case is the non-interacting limit with $g=0$. In this limit, we
have $k_{1}=n_{1}\pi /L$ and $k_{2}=n_{2}\pi /L$, where $n_{1}$ and $n_{2}$
are integers. Since the system is composed of particles with different
masses, the wavefunction can be written as a product state of two
single-particle wavefunctions%
\begin{equation*}
\Psi (x_{1},x_{2})=\phi _{n_{1}}(x_{1})\phi _{n_{2}}(x_{2}),
\end{equation*}%
where
\begin{equation*}
\phi _{n}\left( x\right) =\sqrt{\frac{2}{L}}\sin \frac{n\pi }{L}\left( \frac{%
L}{2}+x\right)
\end{equation*}%
is the eigenstate of the 1D hard-wall potential.

\end{document}